\documentstyle[12pt,epsfig]{article}  
\newcommand{\bea}{\begin{eqnarray}} 
\newcommand{\eea}{\end{eqnarray}} 
\newcommand{\be}{\begin{equation}} 
\newcommand{\ee}{\end{equation}}

\begin{document} 
 
\begin{center} 
{\bfseries Mesons of the rho-family in the P-wave of pion-pion scattering} 
 
\vskip 5mm 
 Yu.S.~Surovtsev$^{a}$ and P.~Byd\v{z}ovsk\'y$^b$ 
 
\vskip 5mm 
{\small 
(a) {\it Bogoliubov Laboratory of Theoretical Physics, JINR, Dubna 141 980, 
Russia } \\ 
(b) {\it Nuclear Physics Institute, Czech Academy of Sciences, 25068 \v{R}e\v{z}, 
Czech Republic }} 
\end{center} 
 
\begin{abstract} 
In the approach, based on analyticity and unitarity and assuming 
an influence of coupled channels, experimental data on the isovector 
$P$-wave of $\pi\pi$ scattering have been analyzed to study $\rho$-like 
mesons below 1.9 GeV. 
\end{abstract}

\section{Introduction} 
 
The investigation of vector mesons is an actual problem up to now due to 
their role in forming {the} electromagnetic structure of particles and 
because, {\it e.g.}, in the $\rho$-family, only the $\rho(770)$ meson can 
be deemed to be well understood \cite{PDG-06}. The other $\rho$-like 
mesons must be either still confirmed in various experiments and analyses 
or their parameters essentially corrected. For example, the $\rho(1250)$ 
meson was discussed actively some time ago \cite{BBS,GG} and it was 
confirmed relatively recently in the amplitude analysis of the LASS 
Collaboration \cite{Aston-LASS} and in combined analysis of several 
processes \cite{Henner}. However this state is referred to only slightly 
in the PDG issue \cite{PDG-06} (the relevant observations are listed 
under the $\rho(1450)$). 
 
On the other hand, the $\pi\pi$ interaction plays a central role in physics 
of strongly interacting particles and, therefore, it has always been an 
object of continuous investigation. Let us note only some recent works 
devoted also to the theoretical study of the isovector $P$-wave of $\pi\pi$ 
scattering. First, there are the analyses of available experimental data on 
the $\pi\pi$ scattering utilizing the Roy equations \cite{Leutwyler,KLL,CCL} 
and the forward dispersion relations \cite{Yndurain,Kaminski}, in which, 
{\it e.g.}, the low-energy parameters of the $\pi\pi$ scattering were 
obtained. Second, there are the works in which the low-energy parameters 
are calculated in chiral theories with the linear realization of chiral 
symmetry \cite{Volkov,BOM}. 
 
We have used our model-independent method \cite{KMS-nc96} based on the 
first principles (analyticity and unitarity) directly applied to analysis 
of experimental data, aiming at studying the $\rho$-like mesons below 1.9 
GeV  and obtaining the $\pi\pi$-scattering length. Unfortunately this method, 
using essentially a uniformizing variable, is applicable only in the 2-channel 
case. Here the $\pi\pi$ and $\omega\pi$ channels are allowed for explicitly 
(in the threshold range of the latter, one has observed a deviation from 
elasticity of the $P$-wave $\pi\pi$ scattering). Influence of other coupled 
channels is supposed to be taken into account through the background. 
In order to investigate the coupling of resonances with these other channels, 
we apply also multichannel Breit--Wigner forms to generate the resonance poles. 
 
The paper is organized as follows. In Section II, we outline the method of 
the uniformizing variable in applying it to studying the 2-channel $\pi\pi$ 
scattering and present results of the analysis of the available data 
\cite{Protopopescu}--\cite{Estabrooks} on the isovector $P$-wave of $\pi\pi$ 
scattering. Section III is devoted to analysis of the same data using the 
Breit--Wigner forms. Finally, in Section VI, we summarize and discuss obtained 
results. 
 
\section{Method of the uniformizing variable} 
 
Let the $\pi\pi$-scattering $S$-matrix be determined on the 4-sheeted 
Riemann surface with the right-hand branch-points at $4m_{\pi^+}^2$ and 
$(m_\omega+m_{\pi^0})^2$ and also with the left-hand one at $s=0$. It is 
supposed that influence of other branch points can be taken into account 
through the background. The Riemann-surface sheets are numbered according 
to the signs of analytic continuations of the channel momenta 
$$k_1=\frac{1}{2}\sqrt{s-4m_{\pi^+}^2} ~~~{\rm and}~~~ 
k_2=\frac{1}{2}\sqrt{s-(m_\omega+m_{\pi^0})^2}$$ as follows: 
$\mbox{signs}({\mbox{Im}}k_1,{\mbox{Im}}k_2)= ++,-+,--,+-$~ 
correspond to sheets I, II, III, IV, respectively. 
 
The $S$-matrix is supposed to be ~$S=S_{res}S_{bg}$ where $S_{res}$ represents 
resonances and $S_{bg}$, the background. In general, an explicit allowance 
for the $(m_\omega+m_{\pi^0})^2$ branch point would permit us to describe 
transitions between the $\pi\pi$ and $\omega\pi$ initial and final states 
with the help of the only one function $d(k_1,k_2)$ ~(the Jost matrix 
determinant) using the Le Couteur--Newton relations \cite{LN}. Unfortunately, 
data on the process $\pi\pi\to\omega\pi$ are absent. 
 
In Ref. \cite{KMS-nc96} it was shown how one can obtain the multichannel 
resonance representations by poles and zeros on the Riemann surface with 
the help of the formulae, expressing analytic continuations of the matrix 
elements, describing the coupled processes, to unphysical sheets in terms of 
those on sheet I. It is convenient to start from resonance zeros on sheet I. 
Then in the 2-channel $\pi\pi$ scattering, we have three types of resonances: 
({\bf a}) described by a pair of complex conjugate zeros in the $S$-matrix 
element on sheet I and by a pair of conjugate shifted zeros on sheet IV; 
({\bf b}) described by a pair of conjugate zeros on sheet III and by a pair 
of conjugate shifted zeros on sheet IV; ({\bf c}) which correspond to a pair 
of conjugate zeros on sheet I, a pair on sheet III and two pairs of conjugate 
zeros on sheet IV. The poles on sheet II, III and IV are situated in the same 
energy points as the corresponding zeros on sheet I, IV and III, respectively. 
Note that the size of shift of zeros on sheet IV relative to the ones on 
sheets I and III is determined by the strength of coupling of the channels 
(here $\pi\pi$ and $\omega\pi$). The cluster kind is related to the nature 
of resonance. 
 
With the help of the uniformizing variable\footnote{The analogous uniformizing 
variable has been used, {\it e.g.}, in Ref. \cite{Meshch} in studying the 
forward elastic $p{\bar p}$ scattering amplitude and in Ref. \cite{SKN-epja02} 
in the combined analysis of data on processes $\pi\pi\to\pi\pi,K\overline{K}$ 
in the channel with $I^GJ^{PC}=0^+0^{++}$.} 
\begin{equation} \label{v} 
v=\frac{(m_\omega+m_{\pi^0})/2~\sqrt{s-4m_{\pi^+}^2} + 
m_{\pi^+}~\sqrt{s-(m_\omega+m_{\pi^0})^2}}{\sqrt{s\left(\left((m_\omega+ 
m_{\pi^0})/2\right)^2-m_{\pi^+}^2\right)}}, 
\end{equation} 
the considered 4-sheeted Riemann surface is mapped onto the $v$-plane, divided 
into two parts by the unit circle centered at the origin. Sheets I (II), III (IV) 
are mapped onto the exterior (interior) of the unit disk on the upper and lower 
$v$-half-plane, respectively. The physical region extends from the point $i$ 
on the imaginary axis ($\pi\pi$ threshold) along the unit circle clockwise in 
the 1st quadrant to the point 1 on the real axis ($\omega\pi^0$ threshold) and 
then along the real axis to the point 
$b=\sqrt{(m_\omega+m_{\pi^0}+2m_{\pi^+})/(m_\omega+m_{\pi^0}-2m_{\pi^+})}$ into 
which $s=\infty$ is mapped on the $v$-plane. The intervals 
$(-\infty,-b],[-b^{-1},b^{-1}],[b,\infty)$ on the real axis are the images 
of the corresponding edges of the left-hand cut of the $\pi\pi$-scattering 
amplitude. The ({\bf a}) resonance is represented in $S(\pi\pi\to\pi\pi)$ 
by two pairs of poles on the images of sheets II and III, symmetric to each 
other with respect to the imaginary axis, and by zeros, symmetric to these 
poles with respect to the unit circle. Note that the symmetry of the zeros 
and poles with respect to the imaginary axis appears due to the real 
analyticity of the $S$-matrix, and the symmetry of the poles and zeros with 
respect to the unit circle ensures a realization of the known experimental 
fact that the $\pi\pi$ interaction is practically elastic up to a vicinity 
of the $\omega\pi^0$ threshold. 
 
The resonance part of $S$-matrix $S_{res}$ becomes a one-valued function on 
the $v$-plane and, in the $\pi\pi$ channel, it is expressed through the 
$d(v)$-function as follows\footnote{Other authors also have used the 
parameterizations with the Jost functions at analyzing the $S$-wave $\pi\pi$ 
scattering in the one-channel approach \cite{Bohacik} and in the two-channel 
one \cite{MP-93}. In latter work, the uniformizing variable $k_2$ has been 
used and the $\pi\pi$-threshold branch-point has been neglected, therefore, 
their approach cannot be employed near by the $\pi\pi$ threshold.}: 
\begin{equation} \label{v:S} 
S_{res}=\frac{d(-v^{-1})}{d(v)} 
\end{equation} 
where $d(v)$ represents the contribution of resonances, described by one 
of three types of the pole clusters in the 2-channel case, {\it i.e.}, 
\begin{equation} \label{d_res} 
d = v^{-M}\prod_{n=1}^{M} (1-v_n^* v)(1+v_n v) 
\end{equation} 
with $M$ the number of pairs of the conjugate zeros. 
 
The background part $S_{bg}$ is taken in the form: 
\begin{equation} \label{S_bg} 
S_{bg}=\exp\left[2i\left(\sqrt{\frac{s-4m_{\pi^+}^2}{s}}\right)^3\left(\alpha_1+ 
\alpha_2~\frac{s-s_1}{s}~\theta(s-s_1)+\alpha_3~\frac{s-s_2}{s}~\theta(s-s_2)
\right) \right] 
\end{equation} 
where $\alpha_i=a_i+ib_i$, $s_1$ is the threshold of 4$\pi$ channel 
noticeable in the $\rho$-like meson decays, $s_2$ is the threshold of 
$\rho2\pi$ channel. Due to allowing for the left-hand branch-point at $s=0$ 
in the uniformizing variable (\ref{v}), $a_1=b_1=0$. Furthermore, $b_2=0$ 
is an experimental fact. 
 
With formulas (\ref{v:S})--(\ref{S_bg}), we have analyzed data 
\cite{Protopopescu}--\cite{Estabrooks} for the inelasticity parameter ($\eta$) 
and phase shift of the $\pi\pi$-scattering amplitude ($\delta$): 
$S(\pi\pi\to\pi\pi)=\eta\exp(2i\delta)$. First, we considered the data of Refs. 
\cite{Protopopescu,Hyams} introducing three ($\rho(770)$, $\rho(1250-1580)$ 
and $\rho(1550-1780)$) and four (the indicated ones plus $\rho(1860-1910)$) 
resonances. 
 
From a variety of possible resonance representations by pole-clusters, 
the analysis selects the following one to be the most relevant: the 
$\rho(770)$ is described by the cluster of type ({\bf a}) and the other 
resonances by type ({\bf b}). Description in both assumed cases is 
satisfactory: $\chi^2/\mbox{NDF}$ is $204.795/(146-15)=1.563$ (three 
resonances) and $197.936/(146-19)=1.558$ (four resonances). The background 
parameters are: $a_2=0.0039$, $a_3=0.0253$, and $b_3=-0.0337$ for the case 
of three resonances and $a_2=-0.00065$, $a_3=0.00432$, and $b_3=0.0001$ 
for four resonances. When calculating $\chi^2$ for the inelasticity parameter, 
three points of data \cite{Hyams} at 990, 1506 and 1825 MeV have been omitted 
in both cases as giving the anomalously big contribution to $\chi^2$. When 
calculating $\chi^2$ for the phase shift, three points of data \cite{Hyams} 
at 1841, 1804 and 1882 MeV have been omitted in the case of three resonances 
and those at 765, 1643 and 1804 MeV in the case of four resonances. 
 
Let us show the obtained pole clusters on the lower $\sqrt{s}$-half-plane 
in Table~\ref{clusters} (it is clear that there are also complex-conjugate 
poles on the upper half-plane). 
\begin{table}[ht] \centering\caption{Pole clusters for the $\rho$-like 
resonances.} 
\vskip0.3truecm 
{ 
\begin{tabular}{|c|c|c|c|c|}\hline\multicolumn{5}{|c|} {Three resonances}\\ 
\hline \multicolumn{2}{|c|}{Sheet} & 
II & III & IV \\ \hline {$\rho(770)$} & {$\sqrt{s_r}$, MeV} & 
$767-i(74.8)$ & $786.5-i(64.1)$ & {} \\ 
\hline {$\rho(1250)$} & {$\sqrt{s_r}$, MeV} & 
{} & $1249.9-i(156.5)$ & $1249.6-i(153.1)$ \\ 
\hline {$\rho(1600)$} & {$\sqrt{s_r}$, MeV} 
& {} & $1576-i(141.7)$ & $1575.6-i(78.4)$ \\ 
\hline 
\multicolumn{5}{|c|} {Four resonances}\\ 
\hline \multicolumn{2}{|c|}{Sheet} & 
II & III & IV \\ \hline {$\rho(770)$} & {$\sqrt{s_r}$, MeV} & 
$767.2-i(74.8)$ & $785.2-i(64.1)$ & {} \\ 
\hline {$\rho(1250)$} & {$\sqrt{s_r}$, MeV} & 
{} & $1249.5-i(156.4)$ & $1249.7-i(152.2)$ \\ 
\hline {$\rho(1600)$} & {$\sqrt{s_r}$, MeV} 
& {} & $1578.7-i(142.5)$ & $1578.2-i(77.1)$ \\ 
\hline {$\rho(1900)$} & {$\sqrt{s_r}$, MeV} & 
{} & $1871.5-i(95.7)$ & $1894.9-i(98.2)$ \\ 
\hline 
\end{tabular}} 
\label{clusters} 
\end{table} 
 
\begin{figure}[ht] 
\begin{center} 
\hspace*{-0.9cm} 
\epsfig{file=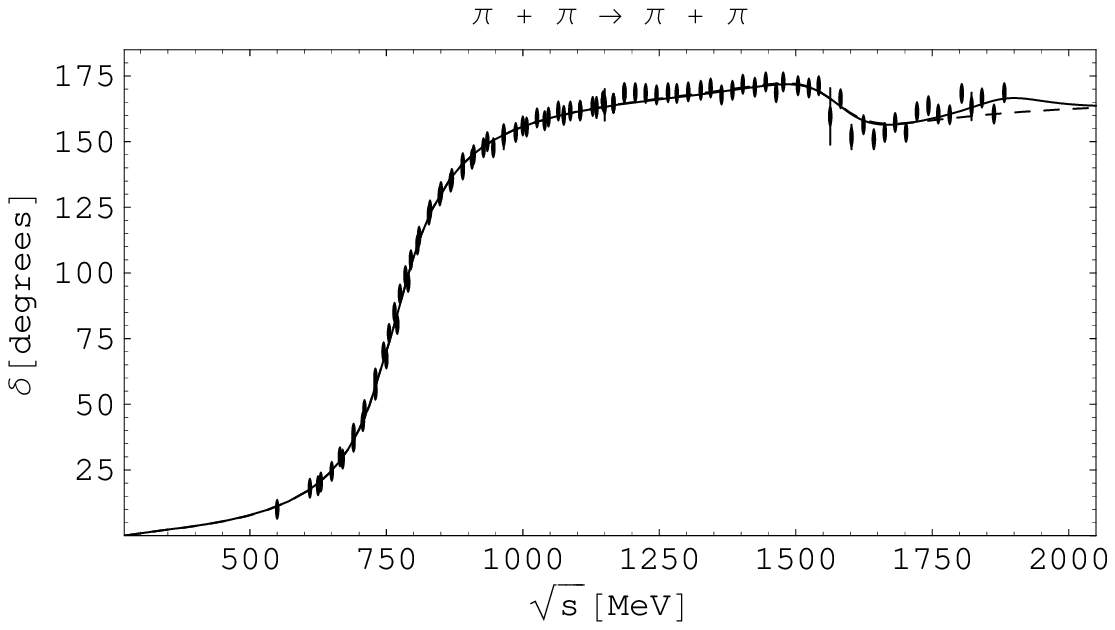,width=9cm} 
\hspace*{-1.1cm} 
\epsfig{file=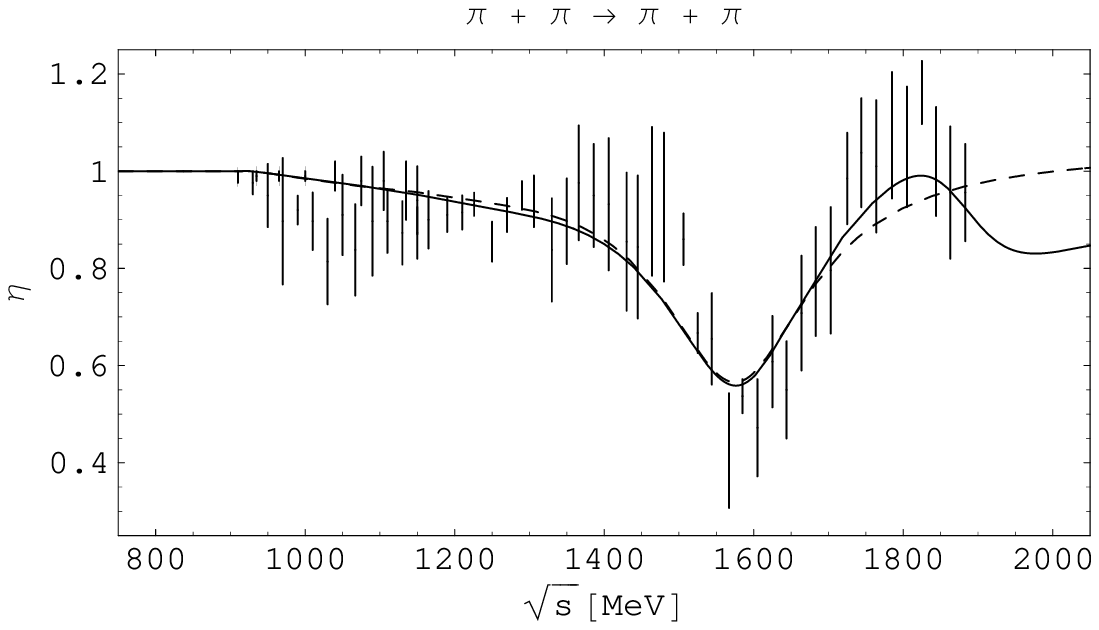,width=9cm} 
\end{center} 
\vskip -.3cm 
\caption{The phase shift of amplitude (the left part) and module of matrix 
element (the right part) of the $P$-wave $\pi\pi$-scattering. The dashed 
curve -- 3-resonance description; the solid curve -- 4-resonance one. 
The data are from Refs. \cite{Protopopescu,Hyams}.} 
\label{fig:m-ind.phs.mdl} 
\end{figure}

On figures \ref{fig:m-ind.phs.mdl}, we demonstrate results from our fitting 
to data \cite{Protopopescu,Hyams}. The dashed curve is for the three-resonance 
description and the solid one for the four-resonance case. 
 
Though $\chi^2/\mbox{NDF}$ is practically the same in both cases, consideration 
of the obtained parameters and energy dependence of the fitted quantities 
suggests that the fourth resonance $\rho(1900)$ is desired. Therefore, in the 
4-resonance picture, we have also carried out an analysis of data \cite{Estabrooks} 
jointly with the already considered data \cite{Protopopescu,Hyams}. 
In Ref.\cite{Estabrooks}, results of two analyses are cited: one uses 
the $s$-channel helicity amplitude when extracting the $\pi\pi$-scattering 
amplitude on the $\pi$-exchange pole; in the other, the $t$-channel one is used 
instead. Therefore, we have taken both analyses as independent. There are given 
the data for the phase shift of amplitude below the $K\overline{K}$ threshold. 
Comparing these data with the ones of Refs.\cite{Protopopescu,Hyams}, one can 
see that the points of the former lie systematically by 1$^0$-5$^0$ higher than 
the ones of the latter, except for two points of Ref.\cite{Protopopescu} at 710 
and 730~MeV, which lie by about 2$^0$ higher than the corresponding points of 
Ref.\cite{Estabrooks} and which are omitted in the subsequent analyses. Since 
we do not know the energy dependence of the remarked deviations of points, we 
have supposed a constant systematic error that must be determined in the 
combined analysis of data. We have obtained a satisfactory description with 
$\chi^2/\mbox{NDF}=284.777/(186-19)=1.705$ and with the indicated systematic 
error equal to $-1.9^0$. 
 
Let us indicate (Table~\ref{v:zeros}) the obtained resonance zero positions 
on the right-hand $v$-half-plane (there are also zeros symmetric to the 
indicated ones with respect to the imaginary axis). 
\begin{table}[ht] \centering\caption{Positions of the resonance zeros on 
the right-hand $v$-half-plane.} 
\vskip0.3truecm 
{\begin{tabular}{|l|ll|}\hline 
\multicolumn{1}{|l|}{Resonance} & \multicolumn{2}{|c|} {$v_n$}\\ 
\hline 
{$\rho(770)$} & $v_1=1.044245+0.215405i$ & $v_2=0.926647-0.178156i$ \\ 
\hline 
{$\rho(1250)$} & $v_3=1.238348-0.032333i$ & $v_4=0.806223-0.021908i$\\ 
\hline 
{$\rho(1600)$} & $v_5=1.282453-0.007478i$ & $v_6=1.282453-0.007478i$ \\ 
\hline 
{$\rho(1900)$} & $v_7=1.305758-0.005171i$ & $v_8=0.766563-0.003215i$ \\ 
\hline 
\end{tabular}} 
\label{v:zeros} 
\end{table} 
 
The background parameters are: $a_2=0.00599$, $a_3=0.02585$ and $b_3=-0.00498$. 
 
\begin{figure}[ht] 
\begin{center} 
\hspace*{-0.9cm} 
\epsfig{file=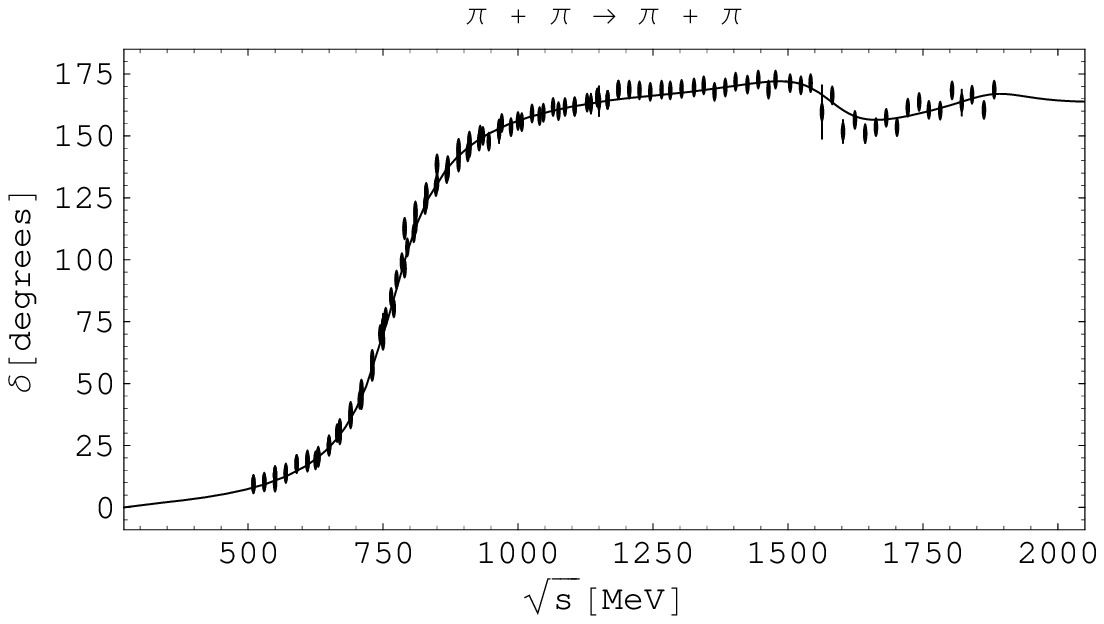,width=9cm} 
\hspace*{-1.1cm} 
\epsfig{file=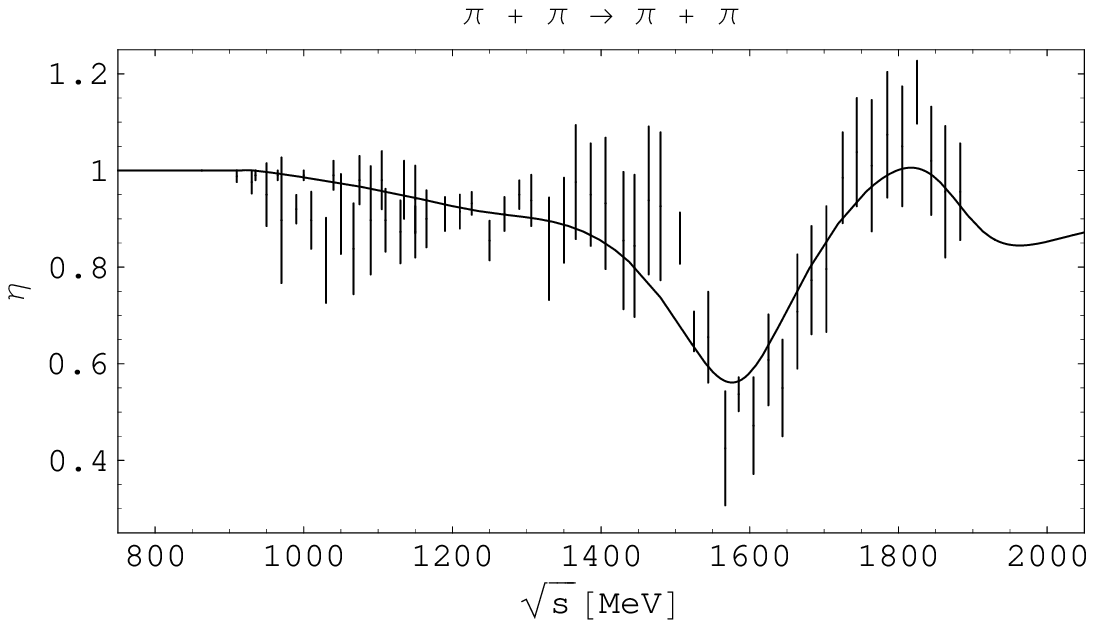,width=9cm} 
\end{center} 
\vskip -.3cm 
\caption{The phase shift of amplitude and module of matrix element of the 
$P$-wave $\pi\pi$-scattering. The data from Refs 
\cite{Protopopescu}-\cite{Estabrooks}.} 
\label{fig:m-ind.phs.mdl2} 
\end{figure}

On figures~\ref{fig:m-ind.phs.mdl2}, we demonstrate energy dependences of the 
analyzed quantities compared with all used experimental data 
\cite{Protopopescu}--\cite{Estabrooks}. 
 
In Table~\ref{clusters2}, we show the obtained pole clusters on the lower 
$\sqrt{s}$-half-plane. 
\begin{table}[ht] \centering\caption{Pole clusters for the $\rho$-like 
resonances.} 
\vskip0.3truecm 
{\begin{tabular}{|c|c|c|c|c|} 
\hline \multicolumn{2}{|c|}{Sheet} & 
II & III & IV \\ \hline {$\rho(770)$} & {$\sqrt{s_r}$, MeV} & 
$767.1-i(73.3)$ & $783-i(65.8)$ & {} \\ 
\hline {$\rho(1250)$} & {$\sqrt{s_r}$, MeV} & 
{} & $1251.2-i(152.2)$ & $1249.5-i(144.6)$ \\ 
\hline {$\rho(1600)$} & {$\sqrt{s_r}$, MeV} 
& {} & $1584.6-i(141.3)$ & $1579.7-i(75)$ \\ 
\hline {$\rho(1900)$} & {$\sqrt{s_r}$, MeV} & 
{} & $1872.3-i(96.6)$ & $1895.5-i(95)$ \\ 
\hline 
\end{tabular}} 
\label{clusters2} 
\end{table} 
 
Masses and widths of the obtained $\rho$-states can be calculated from the pole 
positions on sheet II for resonances of type ({\bf a}) and on sheet IV for 
resonances of type ({\bf b}). If the resonance part of the amplitude reads as 
$$T^{res}=\frac{\sqrt{s}~\Gamma_{el}}{m_{res}^2-s-i\sqrt{s}~\Gamma_{tot}},$$ 
we obtain for the masses and total widths, respectively, the following 
values (in the MeV units): 
\begin{center} 
{\begin{tabular}{lllll} for & ~~$\rho(770)$, & ~~770.6 & ~~and & ~~146.6;\\ 
for & ~~$\rho(1250)$, & ~~1257.8 & ~~and & ~~289.2;\\ 
for & ~~$\rho(1600)$, & ~~1581.5 & ~~and & ~~150;\\ 
for & ~~$\rho(1900)$, & ~~1897.9 & ~~and & ~~190.\end{tabular}}\end{center} 
 
\section{The Breit--Wigner analysis of $P$-wave $\pi\pi$ scattering} 
 
In various works \cite{PDG-06}, it was shown that the $\rho$-like resonances 
obtained in the previous section have also other considerable decay channels 
in addition to those considered explicitly above. It was observed that the 
$\rho(1450)$ (and/or $\rho(1250)$?) have also the decay modes: $\eta\rho^0$ 
($<4\%$), $4\pi$ (seen), and $\phi\pi$ ($<1\%$). The $\rho(1700)$ has also 
the large branching into the $4\pi$ (large), $\rho2\pi$ (dominant), and 
$\eta\rho^0$ (seen) channels. 
 
To include explicitly influence of several interested channels and to obtain 
information about couplings with these channels only on the basis of analysis 
of the $\pi\pi$-scattering data, we have used 5-channel Breit--Wigner forms in 
constructing the Jost matrix determinant $d(k_1,\cdots,k_5)$, to generate the 
resonance poles and zeros in the Le~Couteur--Newton relation: 
\begin{equation} \label{CN:S} 
S_{res}=\frac{d(-k_1,\cdots,k_5)}{d(k_1,\cdots,k_5)}. 
\end{equation} 
In eq.(\ref{CN:S}) $k_1$, $k_2$, $k_3$, $k_4$ and $k_5$ are the $\pi\pi$-, 
$\pi^+\pi^-2\pi^0$-, $2(\pi^+\pi^-)$-, $\eta2\pi$- and $\omega\pi^0$-channel 
momenta, respectively. The $d$-function is taken as ~$d=d_{res}d_{bg}$ with 
$d_{res}$, describing resonance contributions, and $d_{bg}$, the background. 
 
The Breit--Wigner form for the resonance part of the $d$-function is assumed 
as 
\begin{equation} 
d_{res}(s)=\prod_{r} 
\left[M_r^2-s-i\sum_{j=1}^5\rho_{rj}^3~R_{rj}~f_{rj}^2\right], 
\end{equation} 
where $\rho_{rj}=k_j(s)/k_j(M_r^2)$, $f_{rj}^2/M_r$ is the partial width of 
resonance with mass $M_r$, $R_{rj}$ is a Blatt--Weisskopf barrier factor 
\cite{Blatt-Weisskopf} conditioned by the resonance spins. For the vector 
particle this factor have the form: 
\begin{equation} 
R_{rj}=\frac{1+\frac{1}{4}(\sqrt{M_r^2-4m_j^2}~r_{rj})^2} 
{1+\frac{1}{4}(\sqrt{s-4m_j^2}~r_{rj})^2} 
\end{equation} 
with radius $r_{rj}=0.705$ fermi for all resonances in all channels as a result 
of our analysis. Furthermore, we have assumed that the widths of resonance 
decays to $\pi^+\pi^-2\pi^0$ and $2(\pi^+\pi^-)$ channels are related with each 
other by relation: $f_{r2}=f_{r3}/\sqrt{2}$. 
 
The background part $d_{bg}$ is 
\begin{equation} \label{d bg} 
d_{bg}=\exp\left[-i\left(\sqrt{\frac{s-4m_{\pi^0}^2}{s}}\right)^3\left(\alpha_1+ 
\alpha_2~\frac{s-s_1}{s}~\theta(s-s_1)\right) 
\right] 
\end{equation} 
where $\alpha_i=a_i+ib_i$, $s_1$ is the threshold of $\rho2\pi$ channel (it is 
clear that $a_2$ and $b_2$ take into account also influence of other channels 
opened at higher energies than the $\rho2\pi$ threshold); $b_1$ is taken to 
be zero. 
 
With formulas (\ref{CN:S})--(\ref{d bg}) we have carried out the analysis, 
just as in the previous section, both with three resonances and with the four 
ones. We have obtained the same reasonable description in both cases: the 
total $\chi^2/\mbox{NDF}=316.206/(186-17)=1.871$ for three resonances and equals 
$315.254/(186-22)=1.922$ for four resonances. Let us show the obtained values of 
resonance parameters for the second case in Table~\ref{vparameters}. The systematic 
error of data \cite{Estabrooks}, discussed in the previous section, is equal to 
$-2.031^0$ in this case. When calculating $\chi^2$ for the inelasticity parameter, 
three points of data \cite{Hyams} at 990, 1030 and 1825 MeV have been omitted as 
giving the anomalously big contribution to $\chi^2$. When calculating $\chi^2$ 
for the phase shift, three points of data \cite{Estabrooks} have been omitted: 
one at 790 MeV from the $s$-channel analysis, and two at 790 and 850 MeV from 
the $t$-channel one. For the background we find: $a_1=-0.001\pm0.002$, 
$a_2=-0.1065\pm0.012$, and $b_2=0.00715\pm0.016$. 
 
\begin{figure}[ht] 
\begin{center} 
\hspace*{-0.9cm} 
\epsfig{file=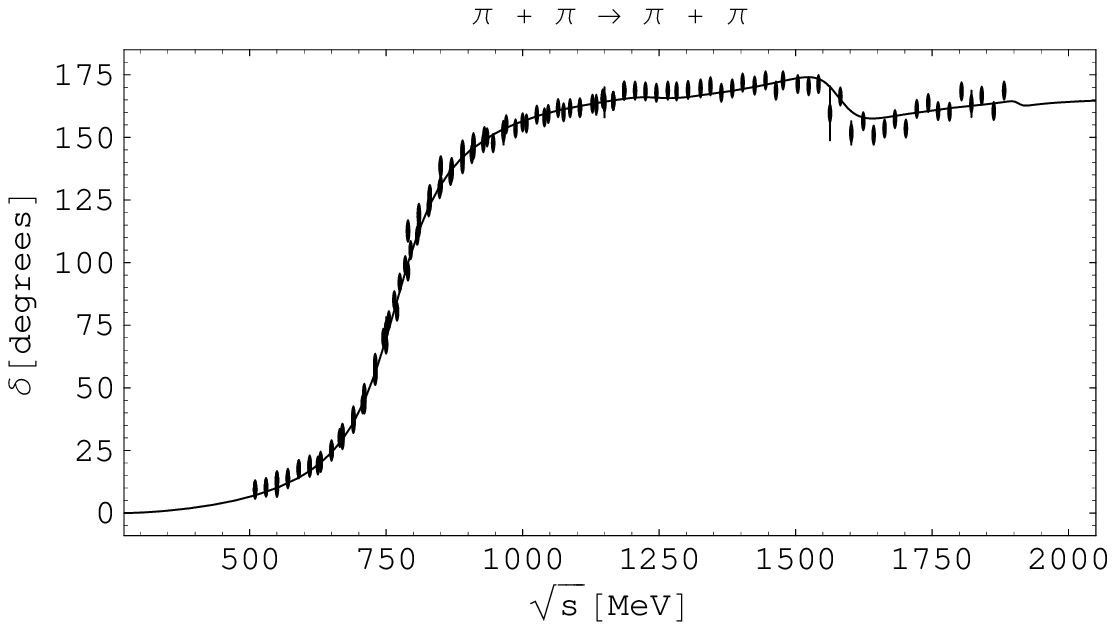,width=9cm} 
\hspace*{-1.1cm} 
\epsfig{file=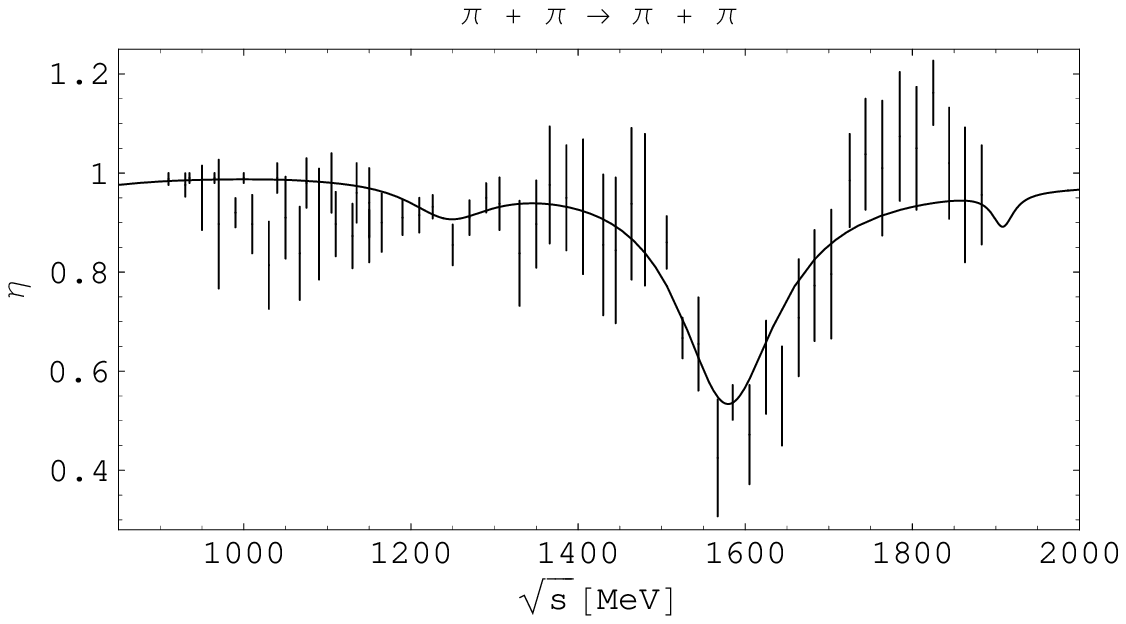,width=9cm} 
\end{center} 
\vskip -.3cm 
\caption{The phase shift of amplitude and module of matrix element of the 
$P$-wave $\pi\pi$-scattering. The curves show result of fitting to data 
\cite{Protopopescu}--\cite{Estabrooks} using the Breit-Wigner form.} 
\label{fig:BWphs.mdl} 
\end{figure}

\begin{table}[ht]\centering\caption{The $\rho$-like resonance parameters (all 
in the MeV units).} 
\vskip0.2truecm 
{\footnotesize 
\begin{tabular}{|c|c|c|c|c|c|c|c|} \hline 
{Resonance} & ~$M$~ & $f_{r1}$ & $f_{r2}$ & $f_{r3}$ & $f_{r4}$ & $f_{r5}$ & 
$\Gamma_{tot}$ \\ 
\hline 
{$\rho(770)$} & 777.57$\pm$0.33 & 343.5$\pm$0.75 & 25.8$\pm$5.9 & 36.5$\pm$8.4 
& {} & {} & $\approx$154.3\\ 
{$\rho(1250)$} & 1247.8$\pm$15.9 & 75.9$\pm$7.5 & 175.8$\pm$40.7 & 248.7$\pm$57.5 & 
176$\pm$112 & 162$\pm$116 & $>$124\\ 
{$\rho(1600)$} & 1580$\pm$4.7 & 247$\pm$5.3 & 236.8$\pm$8.7 & 334.9$\pm$12.3 & 
159$\pm$32 & 106.5$\pm$40.6 & $>$168\\ 
{$\rho(1900)$} & 1908.3$\pm$38.3 & 46.4$\pm$16 & 138 & 195 & 29.7 & 34 & $>$32\\ 
\hline  \end{tabular}} 
\label{vparameters} 
\end{table} 
Note that, in Table \ref{vparameters}, we do not show  the errors for $f_{42}$, 
$f_{43}$, $f_{44}$, and $f_{45}$, because insufficiency of data does not permit 
us to fix reliably these parameters. On figure~\ref{fig:BWphs.mdl}, we demonstrate 
results from our fitting to data \cite{Protopopescu}-\cite{Estabrooks}. 
We may conclude that whereas the model-independent analysis testifies somehow 
in favour of existence of the $\rho(1900)$, the Breit--Wigner approach cannot 
verify this result. 
 
We have also calculated the isovector $P$-wave length of $\pi\pi$ scattering,  
$a_1^1$. Its value is shown in Table~\ref{pipi.length} in comparison with the 
ones known from various evaluations in the local \cite{BOM} and 
non-local \cite{Volkov} Nambu--Jona-Lasinio (NJL) model and from the ones with 
the use of Roy's equations \cite{Yndurain,CCL,KLL} (in the second work one 
apply also the chiral perturbation theory (ChPT)to construct a precise 
$\pi\pi$-scattering amplitude at $s^{1/2}\leq 0.8$ GeV). 
\begin{table}[htb] \centering \caption{Comparison of values of the $\pi\pi$ 
scattering length $a_1^1$ from various approaches.} 
\vskip0.3truecm 
\begin{tabular}{|l|c|l|} \hline $a_1^1[10^{-3}m_{\pi^+}^{-3}]$ & 
{References} & ~~~~~~~~~~Remarks \\ \hline 
\hline $33.98\pm 2.03$ & This paper & Breit--Wigner analysis\\ 
\hline $ 34 $ & \cite{BOM} & Local NJL model \\ 
\hline 
$ 37 $ & \cite{Volkov} & Non-local NJL model  \\ 
\hline 
$37.9\pm 0.5$ & \cite{CCL} & Roy equations using ChPT \\ 
\hline 
$38.4\pm 0.8$ & \cite{Yndurain} & Roy equations \\ 
\hline 
$39.6\pm 2.4$ & \cite{KLL} & Roy equations \\ \hline 
\end{tabular} \label{pipi.length} \end{table} 
 
\section{Conclusions} 
 
The reasonable description of all the accessible experimental data on 
the isovector $P$-wave of $\pi\pi$ scattering for the inelasticity parameter 
($\eta$) and phase shift of amplitude ($\delta$) 
\cite{Protopopescu}-\cite{Estabrooks} have been obtained up to 1.88 GeV based 
on the first principles (analyticity and unitarity) directly applied to 
analysis of the data. Analysis has been carried out in the model-independent 
approach using the uniformizing variable (here the satisfactory description 
is obtained: $\chi^2/\mbox{NDF}=1.558$) and applying multichannel Breit--Wigner 
forms to generate the resonance poles and zeros in the $S$-matrix 
($\chi^2/\mbox{NDF}=1.922$). The aim of analysis (except for obtaining a  
unified formula for the $P$-wave $\pi\pi$ scattering amplitude in the whole 
of investigated energy range) was to study  the $\rho$-like mesons below 1.9 
GeV  and to obtain the $P$-wave $\pi\pi$-scattering length. 
 
For the $\rho(770)$, the obtained value of mass is a little bit smaller in the 
model-independent approach ($770.6$ MeV) and a little bigger in the Breit--Wigner 
($777.57\pm0.33$ MeV) one than the averaged one ($775.5\pm0.4$ MeV) cited 
in the PDG tables \cite{PDG-06}, however, it also occurs in analysis of some 
reactions \cite{PDG-06}. The obtained value of the total width in the first 
case ($146.6$) coincides with the averaged PDG one ($146.4\pm1.1$ MeV) and 
it is a little bit bigger in the second case ($\approx154.3$ MeV) than the 
averaged PDG value, however, this is encountered also in other 
analyses \cite{PDG-06}. Note that predicted widths of the $\rho(770)$ decays 
to the $4\pi$-modes are significantly larger than, {\it e.g.}, the ones 
evaluated in the chiral model of some mesons based on the hidden local 
symmetry added with the anomalous terms \cite{Achasov}. 
 
The second $\rho$-like meson has the mass 1257.8 MeV in the 1st analysis and 
$1247.8\pm15.9$ in the 2nd one. This differs significantly from the mass 
($1459\pm11$) of the 2nd $\rho$-like meson cited in the PDG tables \cite{PDG-06}. 
We told already in Introduction that the $\rho(1250)$ meson was discussed 
actively some time ago \cite{BBS,GG}, and next the evidence for it was 
obtained in some analyses \cite{Aston-LASS,Henner}. To the point, if this 
state is interpreted as the first radial excitation of the $1^+1^{--}$ state, 
then it lies down well on the corresponding linear trajectory with 
a universal slope on the $(n,M^2)$ \cite{Anisovich} (n is the radial quantum 
number of the $q{\bar q}$ state), while the meson with mass $M=1450$ MeV turns 
out to be considerably higher than this trajectory. 
 
The third $\rho$-like meson turns out to have the mass 1580 MeV rather 
than 1720 MeV cited in the PDG tables \cite{PDG-06}. Note that in a number 
of previous analyses of some reactions one has also obtained the resonance 
with mass near 1580 MeV \cite{PDG-06}. However, some time ago it was shown 
that the 1600-MeV region contains in fact two $\rho$-like mesons. This was 
made on the basis of investigation of the consistency of the 2$\pi$ and 
4$\pi$ electromagnetic form factors and the $\pi\pi$-scattering 
length \cite{Erkal} and as a result of combined analysis of data on the 
2$\pi$ and 4$\pi$ final states in the $e^+e^-$ annihilation and 
photoproduction \cite{Donnachie87}. We assume this possibility, {\it i.e.}, 
that in the energy range 1200--1800 MeV there are three 
$\rho$-like mesons, but for the final conclusion the combined analysis of 
several processes with which the investigated resonances are appreciably 
coupled have to be performed. Note also a rather big obtained coupling of 
these $\rho$-like mesons with the 4$\pi$ channels. 
 
Finally, as to the $\rho(1900)$, in this energy region there are practically 
no data on the $P$-wave of $\pi\pi$ scattering. The model-independent analysis, 
maybe, somehow testifies in favour of existence of this state, whereas the 
Breit--Wigner one gives the same description with and without the $\rho(1900)$. 
For more definite conclusion about this state, the $P$-wave $\pi\pi$ 
scattering data above 1.88 GeV are needed. Furthermore, the combined analysis 
of coupled processes should be carried out.\\

Yu.S. acknowledges support provided by the Votruba-Blokhintsev Program for 
Theoretical Physics of the Committee for Cooperation of the Czech Republic 
with JINR, Dubna. P.B. thanks the Grant Agency of the Czech Republic, 
Grant No.202/05/2142.


\begin{thebibliography}{99} 
 
\bibitem{PDG-06} W.-M.~Yao {\em et al.} (PDG), J. Phys. G \textbf{33}, 1 (2006). 
\bibitem{BBS} N.M.~Budnev, V.M.~Budnev, and V.V.~Serebryakov, 
              Phys. Lett. B \textbf{70}, 365 (1977). 
\bibitem{GG} S.B.~Gerasimov and A.B.~Govorkov, Z. Phys. C \textbf{13}, 43 (1982). 
\bibitem{Aston-LASS} D.~Aston {\em et al.}, 
              Nucl. Phis. Proc. Suppl. B\textbf{21}, 105 (1991). 
\bibitem{Henner} T.S.~Belozerova and V.K.~Henner, 
              Phys. Elem. Part. Atom. Nucl. \textbf{29}, part 1, 148 (1998). 
\bibitem{Leutwyler} I.~Caprini, G.~Colangelo, J.~Gasser, and H.~Leutwyler,  
              Phys. Rev. D \textbf{68}, 074006 (2003). 
\bibitem{KLL} R.~Kami{\'n}ski, L.~Le{\'s}niak, and B.~Loiseau, 
            Phys. Lett. B \textbf{551}, 241 (2003). 
\bibitem{CCL} I.~Caprini, G.~Colangelo, and H.~Leutwyler, 
              Int. J. Mod. Phys. A \textbf{21}, 954 (2006). 
\bibitem{Yndurain} J.R.~Pel{\'a}ez and F.J.~Yndur{\'a}in, 
              Phys. Rev. D \textbf{71}, 074016 (2005). 
\bibitem{Kaminski} R.~Kaminski, J.R.~Pelaez, and F.J.~Yndurain, 
              Phys. Rev. D \textbf{74}, 014001 (2006); 
              {\it ibid.}, 079903 (2006), Erratum. 
\bibitem{Volkov} A.A.~Osipov, A.E.~Radzhabov, and M.K.~Volkov, hep-ph/0603130. 
\bibitem{BOM} V.~Bernard, A.A.~Osipov, and U.G.~Meissner, 
              Phys. Lett. B \textbf{285}, 119 (1992). 
\bibitem{KMS-nc96} D.~Krupa, V.A.~Meshcheryakov, and Yu.S.~Surovtsev, 
              Nuovo Cimento A \textbf{109}, 281 (1996). 
\bibitem{Protopopescu} S.D.~Protopopescu {\em et al.}, 
              Phys. Rev. D \textbf{7}, 1279 (1973). 
\bibitem{Hyams} B.~Hyams {\em et al.}, Nucl. Phys. B \textbf{64}, 134 (1973). 
\bibitem{Estabrooks} P.~Estabrooks and A.D.~Martin, 
              Nucl. Phys. B \textbf{79}, 301 (1974). 
\bibitem{LN} K.J.~Le Couteur, Proc. Roy. Soc. A \textbf{256}, 115 (1960); 
             R.G.~Newton, J. Math. Phys. \textbf{2}, 188 (1961). 
\bibitem{Meshch} B.V.~Bykovsky,~V.A. Meshcheryakov, and D.V.~Meshcheryakov, 
             Yad. Fiz. \textbf{53}, 257 (1990). 
\bibitem{SKN-epja02} Yu.S.~Surovtsev, D.~Krupa, and M.~Nagy, 
             Eur. Phys. J. A \textbf{15}, 409 (2002). 
\bibitem{Bohacik} J. Bohacik and H. K\"uhnelt, 
             Phys. Rev. D \textbf{21}, 1342 (1980). 
\bibitem{MP-93} D. Morgan and M.R. Pennington, 
             Phys.~Rev.~D \textbf{48}, 1185 (1993). 
\bibitem{Blatt-Weisskopf} J.~Blatt and V.~Weisskopf, 
         {\it Theoretical nuclear physics}, Wiley, N.Y., 1952. 
\bibitem{Achasov} N.N.~Achasov and A.A.~Kozhevnikov, 
            Phys. Rev. D \textbf{71}, 034015 (2005). 
\bibitem{Anisovich} A.V.~Anisovich, V.V.~Anisovich, and A.V.~Sarantsev, 
            Phys. Rev. D \textbf{62}, 051502 (2000). 
\bibitem{Erkal} C.~Erkal and M.G.~Olsson, Z. Phys. C \textbf{31}, 615 (1986). 
\bibitem{Donnachie87} A.~Donnachie and H.~Mirzaie, 
            Z. Phys. C \textbf{33}, 407 (1987). 
 
 
\end{thebibliography}
\end{document}